\documentclass[article]{aa}
\usepackage{epsfig}
\begin{document} 
\thesaurus{3(08.01.1; 10.01.1; 10.07.3:NGC6397)} 
\title{VLT Observations of  Turnoff stars in the Globular Cluster NGC 6397
\thanks{Based on data collected at Paranal Observatory (ESO, Chile)-Proposal:65.L-0654(A)}}
\author{F. Th\'evenin\inst{1}
\and C. Charbonnel\inst{2}
\and J.A. de Freitas Pacheco\inst{1}
\and T.P. Idiart\inst{3}
\and G. Jasniewicz \inst{4}
\and P. de Laverny \inst{1}
\and B. Plez \inst{4}
}
\offprints{Th\'evenin F.}
\institute{Observatoire de la C\^ote d'Azur, B.P.4229, F-06304 Nice Cedex 4, 
France 
\and
Laboratoire d'Astrophysique de Toulouse, CNRS UMR 5572, 14 avenue 
Edouard Belin, 31400, Toulouse, France
\and
Instituto Astron\^omico e Geof\'isico - Av. Miguel Stefano 4200, 
04301-904 S\~ao 
Paulo, Brazil
\and
UMR 5024, Universit\'e Montpellier II, CC 72, 34095 Montpellier Cedex 5,
France
}
\date{Received date; accepted date} 

\maketitle

\begin{abstract}

VLT-UVES  high resolution  spectra of seven turnoff stars in the metal-poor 
globular
cluster NGC 6397 have been obtained. Atmospheric
parameters and abundances of several elements (Li, Na, Mg, Ca, Sc, Ti, Cr, Fe, 
Ni, Zn and Ba) were derived.
The mean iron abundance is [Fe/H] = -2.02, with no
star-to-star variation.
The mean abundances of  the alpha-elements (Ca, Ti) and of the iron-peak 
elements (Sc, Cr, Ni)
are consistent with abundances derived for field stars of similar metallicity.
Magnesium is also almost solar, consistent with the values found by
Idiart \& Th\'evenin (2000) when non-LTE effects (NLTE hereafter) are taken into 
account. The sodium abundance derived for five stars is essentially solar, but 
one
object (A447) is clearly Na deficient. These 
results are compatible with the expected abundance range estimated from the 
stochastic evolutionary halo model of Argast et al. (2000), in which at the epoch
of [Fe/H] $\sim$ -2 the interstellar medium is supposed to become well-mixed.

\end{abstract}

\section{Introduction}

Stars in globular clusters were traditionally considered to be coeval and 
mono-metallic. However, spectroscopic abundance determinations 
performed in the past several years indicate intra-cluster variations of several 
elements in giant stars for all the studied clusters (see Kraft 1994, Da Costa 
1998 and Sneden 1999 for reviews). 
The most striking, dominant and common abundance variations concern the light 
elements involved in the CNO, NeNa and MgAl proton-capture chains, 
i.e. the CNO, Na, Mg and Al isotopes, 
as well as Li.
The nucleosynthesis responsible for the observed abundance patterns of these  
elements is very well understood, but the site where this nucleosynthesis occurs 
is still 
a matter of debate : did the observed stars pollute their own atmosphere
(the so-called ``evolutionary'', or ``deep mixing'' scenario), or did 
they inherit their peculiar composition at their formation (the so-called 
``primordial'' scenario)? In the later case, were the 
inhomogeneities already present in the interstellar medium from which the 
cluster formed, or did an early generation of relatively massive, rapidly 
evolving stars pollute the intra-cluster gas?

In clusters with abundances derived for a large number of giants, some abundance
patterns appear which are not seen among their field counterparts. In 
particular,
sodium and 
oxygen are clearly anti-correlated. This behaviour was demonstrated to be
present in M3 (Kraft et al. 1992), M5 (Sneden et al. 1992), $\omega$ Cen
 (Paltoglou \& Norris 1989, 
Norris \& da Costa 1995), M13 (Peterson 1980, Kraft et al. 1993, 
1997), M15 (Sneden et  al. 1997),
NGC 7006 (Kraft et al. 1998), M4 (Ivans et al. 1999), 
 M92 (Shetrone 1996a) and M10 (Kraft et al. 1995), 
 over a wide range of metallicity (-2.5$\leq$[Fe/H]$\leq$-1). 
It can easily be understood from a nucleosynthesis point of view : 
$^{23}$Na production via proton capture reactions (NeNa cycle) 
becomes active at the same temperature as the ON branch of the CNO 
bi-cycle 
(T $\sim 3 \times 10^7$K), which is reached in the hydrogen burning shell of 
low mass RGB stars (Denissenkov \& Denissenkova 1990; Langer et al.  1993). 
Thus Na-O anti-correlation can in principle be explained in the framework 
of a deep-mixing scenario along the RGB, while it requires a deeper mixing than 
the C and N anomalies do (Denissenkov \& Weiss 1996, 
Cavallo et al.  1998, Palacios et al. 1999, Weiss et al. 2000).

Aluminium and sodium abundances have been determined only for a few
clusters  and the available data show a clear correlation between these
elements (Shetrone 1996b; Kraft et al. 1997, 1998; Ivans et al. 1999), which are
both anti-correlated with magnesium (Norris \& Da Costa 1995, 
Zucker et al. 1996,
Pilachowski et al. 1996, Sneden et al. 1997, Ivans et al. 1999). 
While the proton capture reaction $^{22}$Ne(p,$\gamma$)$^{23}$Na can explain 
the sodium production in low mass RGB stars, 
the reaction $^{24}$Mg(p,$\gamma$)$^{25}$Al requires a temperature largely 
above that reached  by the hydrogen-burning shell 
(70-85 MK, where the maximum temperature reached in canonical models is only 
of $\sim$ 55 MK). 
According to Ivans et al. (1999), the Al-Mg variations in M4 could be explained 
if aluminium is produced at the expense of the isotopes $^{25}$Mg and $^{26}$Mg, 
but this requires a significant enhancement of the initial abundances of 
these isotopes (Denissenkov et al. 1998).

However, so far, all these abundance variations have been observed only in 
evolved stars 
close to the RGB tip (except in M13 for which very sparse data exist for 
some stars next to the RGB bump). Nevertheless, 
in 47 Tuc, and in NGC 6752 
CN and Na variations are already detected in turnoff 
stars (Cottrell \& Da Costa 1981, Briley et al. 1994, 1995), 
opening the possibility that primordial or external pollution mechanisms 
could also have affected the original chemical pattern 
(see also Denissenkov et al. 1998). 
More recently, from high resolution spectroscopy obtained with the 10m Keck I 
telescope, King et al. (1998) derived LTE abundances for seven subgiant stars 
in the metal-poor cluster M92.
They concluded that the abundance ratios [Cr/Fe], [Ca/Fe] and [Ti/Fe] are the 
same as those observed in the field stars, but remarkable differences exist for 
the ratios [Mg/Fe] and [Na/Fe].
From the theoretical point of view,
 it is difficult to interpret these results in the framework of the 
deep-mixing scenario,
for at least three reasons. First, because the onset of this mechanism 
is supposed to occur only after the star reaches the RGB bump
(e.g. Charbonnel et al. 1998 and references therein). 
Second, because the temperature required to activate the NeNa chain 
is not attained before the star reaches the RGB.
Last but not least, there is no evidence for lithium depletion in any
subgiants.

Observations of less evolved stars down to the turnoff of
globular cluster are badly needed
in order to disentangle the primordial and deep-mixing scenarios.
Here we report high-resolution spectroscopy performed with the VLT
of seven turnoff stars in the globular cluster NGC 6397. This cluster was chosen
because of its relative small distance ($\sim$ 2.2 kpc), which permits the study
of turnoff stars with a reasonable signal-to noise ratio. Castilho
et al. (2000) have recently reported abundances for five giants and eleven
subgiants in this cluster, which are complementary to our own investigation.
The plan of this  paper is the following: in Sect.~2 the observations are 
described
as well as the derivation of the stellar parameters; in Sect.~3 the stellar 
abundances are obtained and results are discussed. Finally, in Sect.~4 
the concluding remarks are given.

\section{Observations and Stellar Parameters}

The program stars  were taken from an extensive list of cluster members  
prepared by Alcaino et al. (1997) and selected, according to their colors,
by their location in the turnoff region of the HR diagram.
The program stars and their basic photometric data are given in Table 1.
Identification numbers are the same as in Alcaino et al. (1997).
The spectra were obtained with the UVES spectrograph (Dekker et al. 2000) 
attached
to the second VLT unit (Kueyen telescope), during June and July 2000. A total of
28 spectra of one hour exposure each were taken, corresponding to four
spectra  per object.
The wavelength range covered by our spectra is 4780--5755 $\rm \AA$ and
5835-6800~$\rm \AA$
and the slit width was about 1.1$''$, corresponding to a spectral resolution of
$\approx$ 40,000.  The mean seeing during the observations was $\sim$ 1$''$.
In order to estimate the S/N ratio of our data, the 
average and the standard
deviation of the relative flux were calculated using bins of 500 pixels,
corresponding to 17$\rm\AA$ and 15$\rm\AA$ for spectra obtained with the
red and the blue spectrograph arms, respectively. The S/N ratio varies from
75 in the blue up to 95 in the red. 

\begin{table*}
\caption[]{Program stars. Photometric data are from Alcaino et al. (1997). 
 Column 7 corresponds to $T_{\rm eff}$ in the scale of Alonso et al. (1996) 
adopting E(B-V)=0.18.
The final adopted temperature (column 10) is defined in the text.}
\begin{flushleft}
\begin{tabular}{llcllllllll}
\hline
$Ident$ & V&S/N(mean) & (B-V) & (V-I)  & $V_{\rm R}(km\, s^{-1}$) & & ${T_{\rm 
eff}}(B-V)$ &
${T_{\rm eff}}(H_{\alpha})$ & ${T_{\rm eff}}(H_{\beta})$&${T_{\rm 
eff}}(adop)$\\
(1)&(2)&(3)&(4)&(5)&(6)&&(7)(K)&(8)(K)&(9)(K)&(10)(K)\\
\hline
A228& 16.068&90 & 0.567& 0.827&14.6& &6199&6172&6185&6185\\
A447& 16.185&80 & 0.584& 0.801&16.7& &6125&6190&6161&6159\\
A575& 16.167&75 & 0.573& 0.813&18.1& &6173&6172&6172&6172\\
A721& 16.203&75 & 0.549& 0.813&15.9& &6279&6280&6280&6280\\
A1406&16.145&75 & 0.560& 0.827&10.1& &6230&6285&6285&6265\\
A2084&16.198&70 & 0.542& 0.830&16.7& &6310&6335&6335&6330\\
A2111&16.011&95 & 0.598& 0.867&24.6& &6066&6080&6075&6075\\
\hline
\end{tabular}
\end{flushleft}
\end{table*}

The magnitude range and the (B-V) colors of the program stars are quite narrow
because we wanted to have similar atmospheric parameters, preventing
differences in the stellar chemical abundances from being due
to their evolutionary stage.

The UVES Data Reduction Standard Pipeline (Ballester et al. 2000) was used for 
the 
reduction of the spectra. Then a detailed data treatment was performed
using  MIDAS and IRAF facilities.
For each star and given spectral range, the four images were first averaged 
and the resulting spectrum was binned by two pixels as well as
corrected to the local standard of rest. At this step, each pixel 
corresponds 
to
$\rm 0.0294\AA$ in the blue and  $\rm 0.0348\AA$ in the red. Finally, each 
image was
normalized with respect to the stellar continuum.
Radial velocities derived from our data are given in Table 1. The mean value
is $\rm 16.7\pm 4.0\, km\,s^{-1}$, consistent with the mean radial velocity of
$\rm 18.9\, km\,s^{-1}$ given by Harris (1996).

The spectral lines selected for  abundance determination purposes are, in 
general, those with accurate log~$gf$  and not blended. Equivalent widths
of  55 lines satisfying our criteria were measured using a Gaussian fitting 
procedure, and are given in Table 2. Errors in equivalent widths were
estimated by using Cayrel's (1987) relation and are typically of the
order of 1.4 m\AA.  If errors in positioning of the continuum are also taken 
into account,
then the expected mean error of our equivalent widths is about 2.8 m\AA.

\begin{table*}
\caption[]{Measured equivalent widths ($\rm m\AA$) and adopted line parameters. 
a: Th\'evenin (1989,1990), b: Wiese et al. (1969, 1980),
c: Warner (1968), d: Miles et al. (1969),
e: Martin et al. (1988), f: Smith et al. (1981), g: Castilho et al. (2000).}
\begin{flushleft}
\begin{tabular}{lllrrrrrrrr}
\hline
Ident&$\lambda (\rm \AA)$&EP(eV)&log~$gf$&   228& 447& 575& 721& 1406& 
2084& 2111\\
\hline
 Zn1& 4810.537& 4.08 & -0.14 c & 9.1  & 8.7  & 9.2  & 7.8  & 0.0  &10.1  & 6.0\\
 Fe1& 4890.763& 2.87 & -0.47 a &32.8  &30.5  &31.7  &30.3  &33.9  &31.8  &35.3\\
 Fe1& 4891.502& 2.85 & -0.17 a &47.7  &41.8  &42.4  &44.0  &43.8  &43.5  &49.5\\
 Cr2& 4824.143& 3.87 & -0.94 a & 8.1  & 6.8  & 5.3  & 5.0  &      &      & 7.5\\
 Fe1& 4903.316& 2.88 & -1.10 a &12.8  &12.0  &12.6  &10.7  &      &      &16.2\\
 Fe1& 4918.998& 2.86 & -0.41 a &33.5  &33.1  &34.8  &29.3  &34.0  &31.7  &39.0\\
 Fe1& 4920.514& 2.83  &0.04 a &55.9  &53.5  &55.3  &49.6  &51.9  &54.3  &59.7\\
 Fe2& 4923.930& 2.89 & -1.43 a &60.9  &60.1  &58.2  &57.3  &55.4  &60.6  &61.4\\
 Ba2& 4934.095& 0.00& -0.16 d &36.3  &29.2  &32.6  &28.4  &25.6  &30.2  &40.5\\
 Fe1& 4966.094& 3.33 & -0.73 a & 9.9  & 8.0  &10.3  & 8.0  &      & 9.9  &    \\
 Ti1& 4981.740& 0.85  &0.52 a &17.5  &25.3  &18.2  &17.5  &18.1  &18.6  &23.9\\
 Ti1& 4991.072& 0.84  &0.41 a &20.0  &20.6  &14.6  &15.7  &16.8  &19.5  &19.8\\
 Ti1& 4999.510& 0.83  &0.25 e &14.4  &16.0  &14.9  & 9.4  &11.4  &14.7  &18.2\\
 Fe1& 5001.870& 3.88& -0.25 a &12.5  &10.9  &10.2  & 9.5  & 9.3  &10.7  &13.2\\
 Fe1& 5014.951& 3.94& -0.28 a &12.5  & 8.2  &10.2  & 9.7  &      & 8.5  & 8.8\\
 Fe2& 5018.446& 2.89& -1.24 a &72.4  &68.3  &68.9  &68.4  &71.4  &67.2  &72.5\\
 Fe1& 5049.827& 2.28& -1.46 a &15.9  &13.1  &17.6  &16.0  &14.9  &14.9  &21.3\\
 Ti2& 5129.162& 1.89& -1.39 e &11.6  &11.8  & 9.2  & 7.9  & 9.2  &12.2  &11.3\\
 Fe1& 5133.699& 4.18  &0.14 a &15.5  &15.6  &15.8  &15.1  &13.1  &10.9  &14.6\\
 Fe1& 5162.281& 4.18  &0.09 a &13.5  &13.5  &11.7  &10.7  &13.9  &10.8  &15.2\\
 Mg1& 5172.698& 2.71& -0.32 a &147.3 &135.3 &123.3 &124.0 &136.5 &137.9 &140.9\\
 Mg1& 5183.619& 2.72& -0.08 a &171.2 &152.8 &143.7 &144.8 &159.9 &153.0 &158.8\\
 Ti2& 5185.908& 1.89& -1.35 e &      &14.5  &12.4  & 8.7  & 8.5  &11.3  &13.8\\
 Fe1& 5192.353& 3.00& -0.50 a &25.9  &27.8  &29.7  &28.9  &25.0  &21.7  &29.1\\
 Fe1& 5194.949& 1.56& -2.15 a &16.2  &12.2  &15.2  &13.3  &15.4  &16.4  &19.7\\
 Cr1& 5206.044& 0.94  &0.03 a &34.8  &32.7  &35.5  &29.9  &33.4  &32.4  &39.0\\
 Fe1& 5216.283& 1.61& -2.17 a &12.7  &12.1  &12.8  &10.5  &11.3  &11.1  &17.7\\
 Fe2& 5234.630& 3.22& -2.31 a &13.5  &10.8  &12.4  & 9.9  &10.9  &      &14.2\\
 Fe1& 5269.550& 0.86& -1.42 a &83.0  &81.4  &81.3  &81.2  &84.1  &84.2  &89.4\\
 Fe2& 5316.620& 3.15& -1.89 a &30.7  &27.2  &32.1  &29.0  &28.5  &29.7  &30.3\\
 Ti2& 5336.794& 1.58& -1.70 e &10.2  &      & 9.2  & 7.2  & 9.8  &      &    \\
 Ca1& 5349.469& 2.71& -0.31 f & 8.0  & 7.9  & 8.4  & 8.3  &      & 8.8  & 8.2\\
 Fe1& 5364.880& 4.44  &0.20 a & 8.9  & 8.0  & 8.7  & 8.5  & 9.2  &      &11.7\\
 Fe1& 5367.476& 4.41  &0.26 a &10.0  &11.5  &10.1  &11.6  &      &10.2  &12.9\\
 Fe1& 5383.380& 4.31  &0.48 a &21.0  &18.0  &21.6  &20.3  &15.6  &20.5  &19.4\\
 Fe1& 5405.785& 0.99& -1.98 a &54.4  &51.5  &51.5  &49.8  &49.5  &47.4  &60.2\\
 Fe1& 5415.210& 4.39  &0.51 a &18.3  &      &18.0  &15.9  &17.7  &15.7  &18.8\\
 Fe1& 5424.080& 4.32  &0.55 a &22.8  &22.5  &24.5  &23.1  &23.9  &19.6  &24.5\\
 Fe1& 5434.534& 1.01& -2.22 a &37.8  &37.9  &41.6  &33.9  &37.1  &36.2  &46.0\\
 Ni1& 5476.921& 1.83& -0.89 e &20.8  &14.6  &19.5  &16.8  &15.5  &13.4  &27.1\\
 Fe1& 5497.526& 1.01& -2.75 a &14.3  &12.0  &15.8  &13.3  &13.9  &11.6  &16.3\\
 Fe1& 5506.791& 0.99& -2.85 a &15.0  &10.8  &10.6  &10.6  &10.6  & 9.9  &17.7\\
 Sc2& 5526.821& 1.77  &0.18 a &10.8  &10.5  &11.4  & 9.5  &11.1  &12.5  &10.6\\
 Mg1& 5528.418& 4.34& -0.47 a &40.7  &39.0  &35.0  &31.8  &38.1  &38.3  &40.3\\
 Ca1& 5588.764& 2.52  &0.21 b &29.5  &30.2  &27.8  &28.1  &26.8  &26.5  &32.8\\
 Ca1& 5601.286& 2.52& -0.52 b & 7.7  &      & 7.8  & 8.0  &      & 7.3  & 6.8\\
 Na1& 5688.217& 2.10& -0.42 g & 6.8  &      & 7.1  &      &      &      & 8.0\\
 Ca1& 5857.459& 2.93&  0.24 b &17.8  &15.8  &17.2  &16.2  &18.3  &18.0  &19.1\\
 Ba2& 6141.727& 0.70& -0.08 d &13.4  &11.0  &12.7  &10.9  &11.2  &12.3  &16.9\\
 Ca1& 6162.180& 1.90& -0.10 a &45.0  &41.4  &43.5  &37.9  &39.6  &41.0  &46.4\\
 Fe1& 6230.736& 2.56& -1.24 a &17.7  &13.6  &15.9  &11.0  &13.6  &12.7  &17.3\\
 Fe1& 6393.612& 2.43& -1.57 a &12.3  &12.9  &13.6  &12.1  &      &11.4  &13.8\\
 Ca1& 6439.083& 2.52&  0.39 a &39.3  &37.5  &37.0  &37.7  &34.1  &36.1  &42.9\\
 Fe1& 6494.994& 2.40& -1.24 a &23.5  &18.1  &20.8  &16.7  &15.8  &18.1  &21.8\\
 Ca1& 6717.687& 2.71& -0.39 a & 8.3  & 8.7  & 8.0  & 8.4  &      &      & 7.9\\
\hline
\end{tabular}
\end{flushleft}
\end{table*}

\subsection{Effective temperatures}

In order to obtain precise photospheric abundances, the effective stellar 
temperature
T$_{\rm eff}$ and the surface gravity g need to be estimated quite accurately. 

The reddening in the direction of NGC 6397 was first estimated by Eggen
(1960),
who derived a color excess E(B-V)=0.15. More recent estimates are those by
Djorgovski (1993) and Alcaino et al. (1997), who obtained respectively E(B-V) = 
0.18 and 0.17. The reddening value is an important correction to the 
photometric data if these are used to derive the effective temperature.
We adopt in the following E(B-V) = 0.18, and as an exercise to check this value
we explored the possibility to use the
NaD interstellar lines. From a de-blending procedure,
we obtained W($\lambda$5889) = 0.301
$\pm$0.065 \AA \, and W($\lambda$5896) = 0.232$\pm$0.050 {\AA} for
the interstellar components.  Using the doublet
ratio method, we found a column density $\rm N(Na)= 2.3\times$10$^{12}$ 
cm$^{-2}$.
The statistical correlation between the color excess and the interstellar
sodium column density derived from data by Cohen (1975) is
\begin{equation}
\rm  E(B-V) = -6.89 + 0.571\,log\,N(Na)
\end{equation}
from which we get E(B-V)=0.17. This result is consistent with previous
independent estimates.

We adopted two different procedures to estimate effective temperatures. First,
temperatures were computed from (B-V)$\rm _o$ colors, using the calibration by
Alonso et al. (1996) for low main sequence stars,  with spectral types ranging 
from F0 to K5 and metallicities in the interval
$\rm -2.5 \leq$[Fe/H]$\leq 0$. This temperature scale was established
from infrared photometry and scaled directly to
$\rm T_{eff}$ determinations via reliable angular
diameter measurements. This color-temperature relationship
is in good agreement with that derived recently by Sekiguchi \& Fukugita (2000) 
for
the same spectral range and metallicities [Fe/H]$\leq -1.5$.
The standard deviation of the relation
(B-V)$\rm _o$ vs T$_{\rm eff}$ is about 130~K, depending slightly on the 
metallicity.
According to Alonso et al. (1996), the mean variation $\rm \Delta T_{\rm 
eff}/\Delta (B-V)$ 
is about 30~K per 0.01\,mag and if we take into account: {\it i)} a typical 
random
photometric error of  0.01\, mag in B and V and {\it ii)} a typical error of 
0.04 mag in
the reddening  (Alcaino et al. 1997),  the expected error in  T$_{\rm eff}$ is 
$\pm$140~K.

\begin{table*}
\caption[]{Atmospheric parameters. log\,$\rm g_{evol}$ was 
calculated for a distance of 2.2 kpc. The ${T_{\rm eff}}(\rm adopted)$ 
corresponds to the mean value of the three determinations reported in Table 1.}
\begin{flushleft}
\begin{tabular}{llllllll}
\hline
$Ident$  & ${T_{\rm eff}}(adop)$ & $\zeta_{turb}({\rm km\,s^{-1}})$ &
log\, $g_{\rm LTE}$ & log $g_{\rm NLTE}$
& log $g_{\rm evol}$&$\rm [Fe/H]_{LTE}$&$\rm [Fe/H]_{NLTE}$\\
\hline
A228& 6185 & 1.30& 3.75& 4.15& 4.05 & -2.25& -2.05\\
A447& 6159 & 1.35& 3.75& 4.15& 4.09 & -2.27& -2.05\\
A575& 6172 & 1.10& 3.75& 4.15& 4.09 & -2.24& -2.03\\
A721& 6280 & 1.20& 3.85& 4.20& 4.13 & -2.23& -2.03\\
A1406&6265 & 1.20& 3.85& 4.20& 4.11 & -2.25& -2.04\\
A2084&6330 & 1.30& 3.90& 4.25& 4.15 & -2.23& -2.03\\
A2111&6075 & 1.15& 3.65& 4.05& 4.00 & -2.24& -2.01\\
\hline
\end{tabular}
\end{flushleft}
\end{table*}

A second independent estimate of the effective temperature can be obtained from
H${\alpha}$ and H${\beta}$ line profiles, since wings are very sensitive 
to this parameter. We have fitted our profiles to
theoretical ones computed with the atmospheric models by Gustafsson et al. 
(1975), taking into account self-resonance broadening according to the
prescriptions by Cayrel \& Traving (1960), and Stark broadening of the
wings (Vidal et al. 1973, Stehl\'e et al. 1983). The code used in our 
computations
is from F.~{ Praderie} $\&$ A. Talavera (private communication).
From our fits, temperatures
can be estimated within errors of about 40K. 
In spite of the difficulty in obtaining good flat-field corrections on
echelle spectra, the use of both H${\alpha}$ and
H${\beta}$ lines reduces errors induced by that procedure, an error 
of 80 K will probably be more realistic. More accurate determinations require
knowledge of the surface gravity and metallicity.

\begin{figure}
\psfig{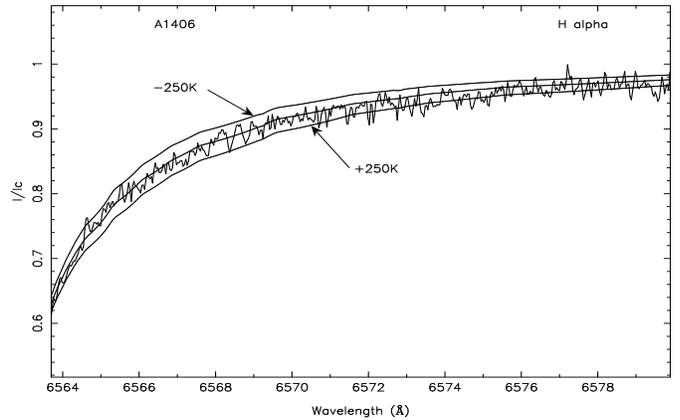}
\caption[]{Fit of $H{\alpha}$ line profile of the star A1406.}
\end{figure}
\begin{figure}
\psfig{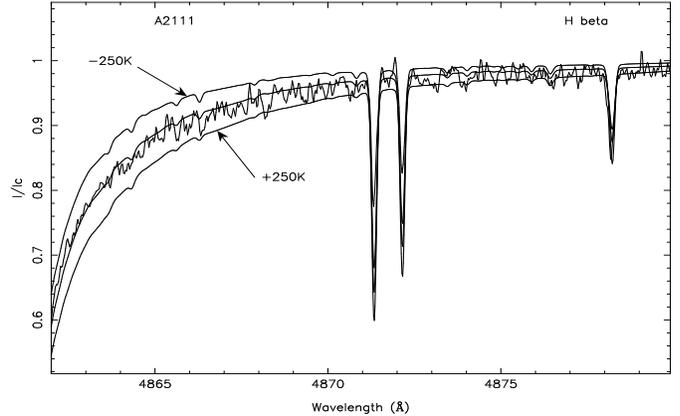}
\caption[]{Fit of $H{\beta}$ line profile of the star A2111.} 
\end{figure}

We therefore used an iterative procedure as well as NLTE gravities
and metallicities derived according to the method  by Th\'evenin \& Idiart 
(1999).
We have verified that, using our procedure and Balmer line profiles, errors
of the order of 0.1 dex in log $g$ induce errors of about 40~K in the
temperature.
A variation of 0.2 dex in the metallicity does not affect noticeably the 
resulting
temperature. In order to illustrate our results, the best fit for two stars 
included in our sample, as well as the resulting profiles
if the temperature is altered by $\pm$250~K are shown in Fig.~1 and Fig.~2.
Temperatures resulting from photometry and fitting of
Balmer lines  are given in Table~1,
as well as the final adopted values (column 10), corresponding 
to the mean of columns 7, 8 and 9. Inspection of these results 
indicate that temperatures derived from both scales are in quite good 
agreement and that no systematic effects are observed.

\subsection{LTE Surface gravities using LTE iron abundances}

The surface gravity is a quantity especially difficult to derive because only
few ionized species have measurable lines in the spectra. Moreover,
in metal-poor dwarfs, reaching ionization balance is difficult
and produces NLTE-overionization
(see Magain 1988; Axer, Furhmann \& Gehren 1995; Th\'evenin \& Idiart 1999).
In a first approximation, adopting LTE,
the surface gravity was estimated by enforcing the equality of abundances 
derived
from FeI and FeII lines, using the curve of growth technique. The 
microturbulent velocity $\rm \zeta_{turb}$
was derived from the plateau of the curve of growth of FeI.
The uncertainties in log~$g$  are due to the fit quality of the curve
of growth for both FeI and FeII, errors in the oscillator strengths  and errors
in the equivalent widths. We expect that total errors in LTE gravities are
not larger than 0.2 dex.

\subsection{NLTE corrections to surface gravities}

Inspection of Table~3 indicates that surface gravities of our program stars 
derived 
under the assumption of LTE correspond to values expected for 
subgiant stars (log $g_{\rm LTE} \sim$ 3.7),
which are incompatible with the actual position of these objects in the HR 
diagram. These low derived gravities are caused by overionization effects in FeI and
they should be corrected according to the procedure
by Th\'evenin \& Idiart (1999) and Idiart \& Th\'evenin (2000). The
strengthening of the UV radiation field produced by a reduced opacity in
metal-poor stars is the main cause of the overionization effect.
Corrected gravities are given in column 5 of Table~3.
NLTE surface gravities can be checked by comparison with those derived 
from the expected value, according to the  position of the star in the HR 
diagram, 
given by the relation
\begin{eqnarray}
\rm log {\it g}_{ evol} = -10.537 + log\,({\it M/M_{\odot}}) + 4\,log {\it 
T}_{\rm eff} + 
\nonumber\\ 
 0.4(V_0 + BC) - 2\,{\rm log}\,d
\end{eqnarray}
where M is the stellar mass, BC is the bolometric correction, d is the distance
in parsecs and the solar quantities log~$g_{\odot}$ = 4.44, 
{\it T}$\rm _{\odot}$ = 5772~K and M$\rm _{bol}(\odot)$ = 4.75 were adopted to 
obtain 
the
numerical constant.
Assuming M = 0.85 M$\rm _{\odot}$ as a typical value for a cluster turnoff star 
and
d = 2200 pc (Djorgovski 1993), the evolutionary gravities can be estimated using 
photometric data and our derived temperatures. These gravities are also given in 
Table~3 (column 6) and are entirely consistent with our NLTE values.

\begin{table*}
\caption[]{LTE abundances with respect to the Sun.}
\begin{flushleft}
\begin{tabular}{lrrrrrrrr}
\hline
$\rm Ident$&  A228&  A447&  A575&  A721& A1406& A2084& A2111&Sun\\
\hline
$\rm [Na/H]$ & -1.92&      & -1.91&      &      &      & -1.83&6.28\\
$\rm [Mg/H]$ & -2.08& -2.17& -2.21& -2.24& -2.11& -2.09& -2.15&7.53\\
$\rm [Ca/H]$ & -1.99& -2.02& -1.99& -1.98& -2.02& -1.99& -1.99&6.36\\
$\rm [Sc/H]$ & -2.00& -2.03& -1.96& -1.98& -1.92& -1.84& -2.07&2.99\\
$\rm [Ti/H]$ & -1.79& -1.69& -1.82& -1.82& -1.78& -1.68& -1.79&4.88\\
$\rm [Cr/H]$ & -2.27& -2.38& -2.36& -2.40& -2.39& -2.38& -2.35&5.61\\
$\rm [Fe/H]$ & -2.25& -2.27& -2.24& -2.23& -2.25& -2.23& -2.24&7.46\\
$\rm [Ni/H]$ & -2.22& -2.42& -2.26& -2.25& -2.31& -2.34& -2.16&6.18\\
$\rm [Zn/H]$ & -1.96& -2.00& -1.96& -1.98&      & -1.84& -2.21&4.60\\
$\rm [Ba/H]$ & -2.45& -2.61& -2.50& -2.51& -2.54& -2.40& -2.41&2.18\\
\hline
\end{tabular}
\end{flushleft}
\end{table*}

\begin{table*}
\caption[]{LTE abundances with respect 
to the Sun (except for Li) using NLTE atmospheric parameters
(e.g. [Fe/H]$\rm _{NLTE}$ and log $g_{\rm NLTE}$, see text).}
\begin{flushleft}
\begin{tabular}{lrrrrrrrr}
\hline
$\rm Ident$&  A228&  A447&  A575&  A721& A1406& A2084& A2111&Sun\\
\hline
$\rm A(Li)$ &   2.22&  2.19&  2.18&  2.21&  2.29&  2.38&  2.22&3.31\\
$\rm [Na1/H]$&  -1.88&      & -1.86&      &      &      & -1.85&6.28\\
$\rm [Mg1/H]$&  -2.13& -2.24& -2.31& -2.27& -2.15& -2.13& -2.26&7.53\\
$\rm [Ca1/H]$&  -1.92& -1.95& -1.93& -1.90& -1.93& -1.88& -1.95&6.36\\
$\rm [Sc2/H]$&  -1.81& -1.83& -1.79& -1.83& -1.75& -1.66& -1.87&2.99\\
$\rm [Ti1/H]$&  -1.72& -1.65& -1.76& -1.76& -1.71& -1.59& -1.71&4.88\\
$\rm [Ti2/H]$&  -1.60& -1.54& -1.65& -1.73& -1.66& -1.51& -1.61&4.88\\
$\rm [Cr1/H]$&  -2.39& -2.46& -2.38& -2.42& -2.34& -2.32& -2.40&5.61\\
$\rm [Cr2/H]$&  -1.97& -2.06& -2.17& -2.18&      &      & -2.03&5.61\\
$\rm [Fe1/H]$&  -2.19& -2.26& -2.20& -2.19& -2.18& -2.16& -2.19&7.46\\
$\rm [Fe2/H]$&  -2.03& -2.11& -2.06& -2.09& -2.07& -2.02& -2.05&7.46\\
$\rm [Ni1/H]$&  -2.16& -2.37& -2.20& -2.21& -2.25& -2.29& -2.09&6.18\\
$\rm [Zn1/H]$&  -1.89& -1.92& -1.88& -1.92&      & -1.76& -2.14&4.60\\
$\rm [Ba2/H]$&  -2.30& -2.45& -2.36& -2.38& -2.41& -2.28& -2.27&2.18\\
\hline
\end{tabular}
\end{flushleft}
\end{table*}

\begin{table*}
\caption[]{NLTE abundances, (:) indicates resonant lines were used}
\begin{flushleft}
\begin{tabular}{lrrrrrrr}
\hline
$\rm Ident$  & A228&  A447&  A575&  A721& A1406& A2084& A2111\\
\hline
$\rm [Na/H]$  &-2.04 &-2.50: &-2.05 &-2.05: &      &-2.15: &-1.99\\
$\rm [Ca/H]$  &-1.82 &-1.79 &-1.81 &-1.78 &-1.80 &-1.80 &-1.75\\
$\rm [Mg/H]$  &-1.96 &-1.99 &-2.05 &-2.08 &-1.96 &-1.94 &-2.00\\
$\rm [Fe/H]$  &-2.05 &-2.05 &-2.03 &-2.03 &-2.04 &-2.03 &-2.01\\
\hline
\hline
$\rm [Na/Fe]$  & 0.01& -0.45:&  0.00& -0.02:&      & -0.12:&  0.02\\
$\rm [Ca/Fe]$  & 0.23&  0.26&  0.22&  0.25&  0.24&  0.23&  0.26\\
$\rm [Mg/Fe]$  & 0.09&  0.06& -0.02& -0.05&  0.08&  0.09&  0.01\\
\hline
\end{tabular}
\end{flushleft}
\end{table*}

\section{The Chemical Abundances}

\subsection{LTE abundances}

LTE abundances were derived by adopting effective temperatures and
gravities given in Table~3 and by using the 
stellar atmosphere models of Gustafsson et al. (1975) and 
recent models issued from the MARCS code (Plez et al. 1992, Asplund et 
al. 1997 and Edvardsson et al. 1993). The resulting
abundances with respect to the solar value (Holweger 1979) and references for 
the adopted log~$\rm gf$ are given in Table~4. For the elements Na, Zn, Sc and Ni 
there was only one faint measurable line and thus the abundance determination for these 
elements is more uncertain when compared to others.
 Typical errors in $\rm T_{\rm eff}$, 
log~$g$, [Fe/H] and $\zeta_{\rm turb}$ are respectively 
equal to 80 K, 0.1 dex, 0.1 dex and 0.2 $\rm km\,s^{-1}$, 
corresponding to abundance errors of about 0.1 dex, which is adopted
here as the error in our final abundance values.

\subsection{LTE abundances deduced using $log\,g_{NLTE}$.}

We have also computed LTE abundances using model atmospheres adopting 
the corrected NLTE iron abundances
and surface gravities (see Table~5), for two main reasons. First, because
the Li abundance is practically unaffected by NLTE effects (Carlsson et al., 
1994).
Thus, LTE values for this element, computed with stellar atmosphere parameters
corrected by overionization effects should give correct abundances.
The lithium abundance was estimated
for all stars included in our sample by fitting synthetic spectra to the 
observed profiles (see, for instance, Jasniewicz et al. 1999). Secondly,
gravities from Hipparcos and Fe II lines alone can be used to overcome the 
ionization
balance difficulty (Th\'evenin \& Idiart 1999).
According to our
computations, the differences between Fe II abundances derived 
with corrected atmospheric parameters (Table~5) and [Fe/H]$_{\rm NLTE}$ are 
small.
This approach
was also followed by Israelian et al. (2001), who studied oxygen and iron in a
sample of metal-poor subdwarfs. They concluded that gravities from Hipparcos are
in good agreement with those estimated in NLTE, reducing the discrepancies 
between oxygen abundances derived from different lines in past 
studies.

\subsection{NLTE abundances}

Because the UV radiation field is increased in metal-poor stars,
the Saha equations are not perfectly satisfied for most
of the studied atoms, which are overionized. Departures from the Boltzmann equations also
exist and in consequence,
NLTE abundances have to be computed using the code by Carlsson (1986) and
atom models for Ca, Mg and Fe according  to Th\'evenin \& Idiart (1999), Idiart \& Th\'evenin 
(2000), and also for Na (Th\'evenin, to be published) for which 
accurate photoionization cross sections are available. This consists of solving for
the statistical equilibrium of a multi-level atom in the stellar atmosphere.
A line by line comparison between the computed equivalent width and the 
observed one, following an iterative procedure, is performed until 
the desired accuracy is attained.
The resulting abundances are given in Table~6.
NLTE abundances of Fe were derived following Th\'evenin \& Idiart (1999), 
i.e.
using the curve of growth technique. $\rm [Fe/H]_{NLTE}$ values
are close to those of [FeII/H](see Table 5), as one should expect. 
The NLTE abundances of Ca, Mg and Na were calculated using the following
transitions: Ca$\lambda$5588, Mg$\lambda$5528 and Na$\lambda$5688 and NaD.
For the Mg$\lambda$5528 line,
mean corrections are $\approx$0.12 dex, in perfect agreement with results 
obtained by
Zhao \& Gehren (2000). Note that the mean NLTE iron abundance and
the mean NLTE Ca abundance give a resulting ratio
[Ca/Fe]$_{\rm NLTE}$ very similar to $\rm [Ca/Fe]_{\rm LTE}$ while
the ratio [Mg/Fe] is slightly more affected by 
NLTE corrections, because NLTE corrections for Mg are significantly lower than those
for Fe, which is not the case for Ca. NLTE corrections strongly depend on the nature of the atom
and of the line transition studied.

It was not possible to measure the equivalent width of the Na$\lambda$5688 line
for all the stars; in this case the resonant doublet was used. Na abundances
derived from the resonant transition are identified with a double dot in
Table~6.

\begin{figure}
\psfig{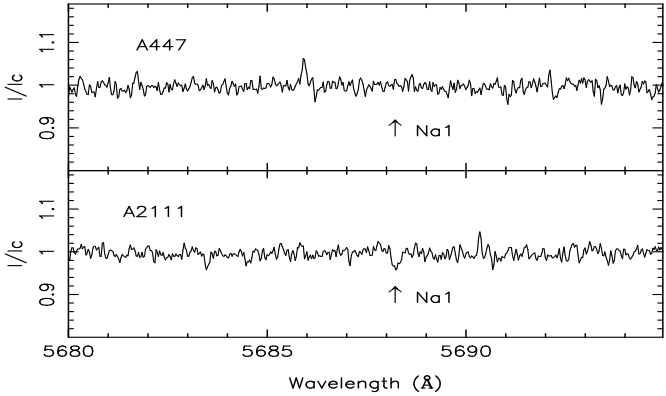}
\caption[]{Spectral region including the Na1(5688 $\rm \AA$) transition for A447 
and
A2111.}
\end{figure}

\section{Discussion}

\subsection{[Fe/H] abundances (NLTE)}

Concerning the resulting [Fe/H] abundances (last column of table 3),
the first point to be emphasized is the rather small star-to-star variation.
The mean NLTE iron abundance is [Fe/H]= -2.02, with a rmsd of only 0.01 dex, to
be compared with our estimated error of 0.10 dex due to the uncertainties in
the effective temperature and surface gravity. Such
a mean value is in good agreement with the iron abundance of giants derived 
by Minniti et al. (1993), namely [Fe/H]= -1.94$\pm$0.04, and with the results by 
Castilho et al. (2000), who performed a study of 16 giants and subgiants in 
this cluster ([Fe/H] = -2.00$\pm$0.05). Five turnoff stars and three subgiants
in NGC 6397 were recently studied by Gratton et al. (2000); they derived a mean
iron abundance in quite agreement with our determination, but they 
emphasize
that the ionization equilibrium is not well reproduced, with abundances from FeI
lines being 0.11 dex higher than those derived from FeII lines. This is inverted
with respect to our expectations we should expect, but the mean temperature adopted by those 
authors
is about 260 K higher than the mean temperature we derived for our turnoff 
stars, and
a similar difference is observed between the adopted mean temperature for 
their
subgiants and those of Castilho et al. (2000).
 
\subsection{$\alpha$-elements: Ca (NLTE), Ti (LTE)}

The mean abundances of calcium and titanium (magnesium will be discussed
separately) from our sample are [Ca/Fe] = +0.24$\pm$0.02 and [Ti/Fe]= 
+0.47$\pm$0.06,
where the former corresponds to NLTE and the latter to LTE calculations.
Again we note a small star-to-star scatter. These results are also
in agreement with LTE abundances derived for evolved stars by Castilho et al. 
(2000), who
obtained [Ca/Fe] = +0.20$\pm$0.07 and [Ti/Fe] = +0.43$\pm$0.12.

According to the results of Idiart \& Th\'evenin (2000), the data points
in the plots [Ca/Fe] or [Mg/Fe] against [Fe/H] for field dwarfs show a ''lumpy''
distribution, which appears, particularly in the case of calcium, when NLTE
effects are taken into account. 
The $\rm [Ca/Fe]_{NETL}$ ratios of our program stars 
match quite well one of those structures, represented as shaded zones
in Figs.~4 and 5 . The reason(s) for the appearance of these structures
is (are) still unknown.

\begin{figure}
\psfig{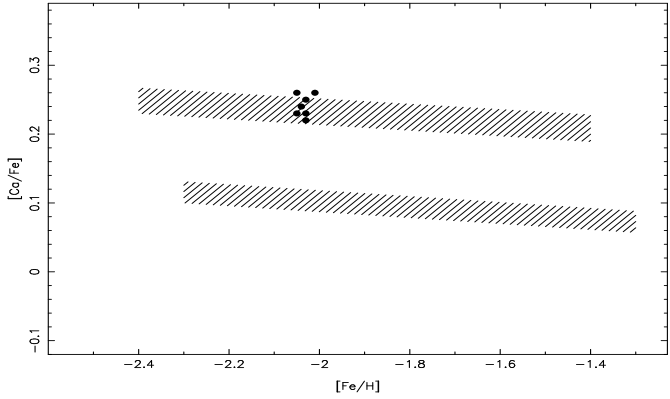}
\caption[]{[Ca/Fe] diagram, black points are NGC6397 results and the shaded
zones are parallel structures from Idiart \& Th\'evenin (2000), see Sect. 4.2, 
adopted error bars are $\pm$0.1 dex for both axes.}
\end{figure}

\begin{figure}
\psfig{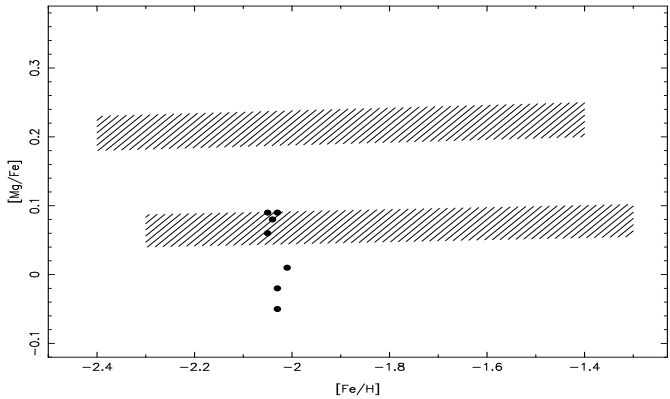}
\caption[]{[Mg/Fe] diagram, same as Fig.~4.}
\end{figure}

\subsection{Iron group (LTE): Sc, Cr, Ni}

Abundances of Sc, Cr and Ni were all derived from LTE calculations. Their
mean values derived from Table 4 are [Sc/Fe] = +0.27$\pm$0.07, [Cr/Fe] = 
-0.12$\pm$0.05,
[Ni/Fe] = -0.03$\pm$0.08. These abundances are entirely compatible with those
derived for field stars of same metallicity (McWilliam et al. 1995; Ryan et al. 
1996).

\subsection{n-capture element: Ba (LTE)}

The only n-capture element studied is barium because of the lack 
of measurable lines for other n-elements.
For field stars, the [Ba/Fe] ratio
increases from values around {-1.5} at [Fe/H]$\approx$-3.5 up to near-solar
values at metallicities around [Fe/H]$\approx$-1.6, and then remains more or 
less
constant. Thus,  NGC 6397 is in a ''transition region''. The mean LTE abundance 
derived from data in Table~4 is [Ba/Fe] = -0.24$\pm$0.06. This value is 
compatible 
with the mean abundance derived from the sample of evolved stars by Castilho 
et al. (2000), namely, [Ba/Fe] = -0.16$\pm$0.12 and is in remarkable good agreement with
abundances of field stars having comparable metallicity (McWilliam et al. 1995;
Ryan et al. 1996).

\subsection{Zinc}

Few results exist on the abundance of Zn in metal poor stars, leaving incomplete
the understanding of
the nucleosynthetic origin of this element. The present status on the Zn abundance
in halo stars was reviewed by Sneden et al. (1991), who showed
that in the metallicity range $-2.5<[Fe/H]<-1.5$ the relative Zn abundance is
$-0.20<[Zn/Fe]<0.25$ dex. Our LTE results (Table 4)
give a mean value $\rm [Zn/Fe]= +0.25$ dex. 
It is worth mentioning that Sneden et al. (1991) used 7.60 dex (see their Table 3 and 4)
for the solar iron abundance, instead of 7.46 adopted in the present work. An additional
correction of -0.02 dex should be applied for Zn due to same reason and there
is a difference of -0.03 between the adopted log gf. Therefore, in order
to compare the results by Sneden et al. (1991) with our mean value, one
should shift the former by $\sim$ 0.1 dex. Under these conditions, our
results are in good agreement with those authors.

Zinc is an interesting element also because its interstellar lines have 
been detected in absorption spectra of quasars (damped Lyman-alpha systems). 
Moreover, zinc can be considered as a fair and useful indicator of the gas 
phase metallicity of interlopers, since it is expected to be far less depleted 
in dust grains than iron.

The [Zn/H] vs redshift correlation was recently reviewed by Savaglio (2000), who
derived the relation
\begin{eqnarray}
\rm [Zn/H] = -0.32z -0.40
\end{eqnarray}
If  one assumes that the gas seen in absorption is associated with the
early ''halo-phase'' of the interlopers and that the
outer galactic halo evolved chemically in a 
similar 
way, then from the
above equation and our derived Zn abundance, one obtains an equivalent 
redshift z $\approx$ 4.75.
This redshift corresponds to an age given by the equation
\begin{eqnarray}
\rm t_{age}(z) = {{1}\over{H_o}}\int_0^z {{dx}\over{(1 + x)\lbrack \Omega_v + 
\Omega_m(1 + x)^3\rbrack^{1/2}}}
\end{eqnarray}
where H$_o$ is the Hubble parameter and we assumed a ``flat'' world model such
as $\rm \Omega_m + \Omega_v$ = 1, where $\rm \Omega_m$ is the density parameter
including both
baryonic and dark matter, and $\Omega_v$ is the density parameter of the vacuum
energy density. Assuming H$\rm _o = 65 km\,s^{-1}Mpc^{-1}$, $\rm \Omega_m$ = 
0.25 and
$\rm \Omega_v$ = 0.75, satisfying of the constraints imposed by primordial 
nucleosynthesis,
type Ia supernovae, Boomerang and Maxima-I data, one obtains t = 14 Gyr for
the presumed age of the cluster. This result must be taken with a grain
of salt since the zinc abundance vs redshift relationship above
is questionable and will probably
not be applicable to the inner metal-rich clusters. These objects have a
higher mean rotational velocity, a smaller velocity dispersion (Zinn 1985;
Borges \& de Freitas Pacheco 1988), and were formed in regions with a higher 
star formation rate.

\subsection{Magnesium (NLTE)}

The mean NLTE abundance of magnesium, derived from Table~6
is [Mg/Fe] = +0.04$\pm$0.06. The NLTE analysis performed by Idiart \&
Th\'evenin (2000) for 252 dwarfs and subgiants indicates a large scatter in
the diagram [Mg/Fe] vs [Fe/H] with some lumpy structures.
Our sample stars are located in the lower envelope of
the diagram by Idiart \& Th\'evenin (see Fig.~5), with four stars on one of those
structures. Similar abundances were 
derived by Gratton et al. (2000), who obtained a mean abundance ratio [Mg/Fe] = 
+0.08$\pm$0.03 from five dwarfs in NGC 6397. Such an agreement remains true
even if their abundances are corrected to our temperature scale.
 
The chemical enrichment of the halo depends on the
type II supernovae rate and on the time-scale required for the ejecta to be
completely mixed with the interstellar gas. Thus, it is reasonable to imagine 
that
the halo was chemically inhomogeneous before total mixing occurs, and that
stars formed during this early phase may have intrinsic abundance differences
at the same metallicity. Model calculations by Argast et al. (2000) (see also
Travaglio et al. 1999) suggest that the halo is unmixed if [Fe/H]$<$-3.0 and 
only when [Fe/H]$>$-2.0 is the halo gas well mixed, with the abundance pattern 
reflecting
the yields integrated over the initial mass function. In the metallicity range
-3.0$<$[Fe/H]$<$-2.0 there is a transition region from the unmixed to the
well-mixed interstellar medium. Field stars can be found in all these phases, 
but
metal-poor globular clusters were mainly formed during the transition period whereas
metal-rich ones were formed when the gas was already well mixed. The large
scatter found in the diagram [Mg/Fe] vs. [Fe/H] by Idiart \& Th\'evenin (2000)
is consistent with the stochastic scenario developed by Argast et al. (2000).

\subsection{Sodium (NLTE)}

The analysis of the sodium abundance in NGC 6397 is more complex. The mean 
abundance
ratio excluding the star A447 is almost solar, namely [Na/Fe]=-0.02$\pm$0.06. 
The
star A447 seems to be sodium-deficient. Fig.~3 shows the spectral region 
including
the transition Na$\lambda$5688 for A447 and A2111. The line is clearly seen in 
the latter but not in the former and this behaviour can be translated into 
abundance
differences, since both stars have similar atmospheric parameters. Note 
that the Na$\lambda$5688 line is difficult to measure even for the star A2111,
and the Na$\lambda$5682 line is undetectable in all spectra because its 
log gf is 0.3 lower when compared to that of the line Na$\lambda$5688. In consequence,
this line could not be extracted from our data since
its equivalent width is always less than 3-4 m\AA. The 
de-blending
of the NaD lines also suggests a sodium deficiency in A447, since the equivalent
width measured for this star is almost one half of that obtained for the 
other objects. This is confirmed because NaD lines also have about one half
the equivalent widths of other stars.
Gratton et al.
(2000) have also derived Na abundances for five dwarfs in NGC 6397. Three stars 
have
almost solar values whereas the other two dwarfs are overabundant. Using our
temperature scale, the mean sodium abundance from their data is
[Na/Fe]=+0.04$\pm$0.11, a value
consistent with our results. Concerning Na abundances in evolved stars, Minniti
et al. (1996) studied five giants in this cluster. Their data indicate a mean 
value
essentially solar but with star-to-star variations. Minniti et al. (1996) 
concluded
that their data are consistent with the Na vs O anti-correlation. However, if 
the 
giants studied by Castilho et al. (2000) as well as those by Gratton et al. 
(2000)
are included in a plot, one obtains essentially a scatter diagram, supporting Na 
abundance variations among evolved stars.

\subsection{Lithium (LTE)}

Lithium is of particular interest since it is of primordial origin and
easily depleted in stellar interiors.

\begin{figure}
\psfig{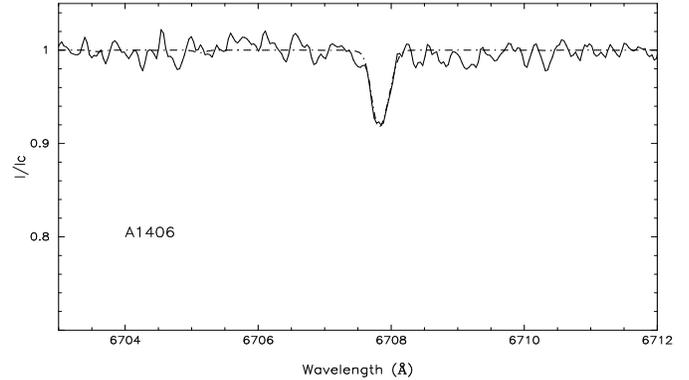}
\caption[]{Fit of the lithium blend for the star A1406.}
\end{figure}

\begin{figure}
\psfig{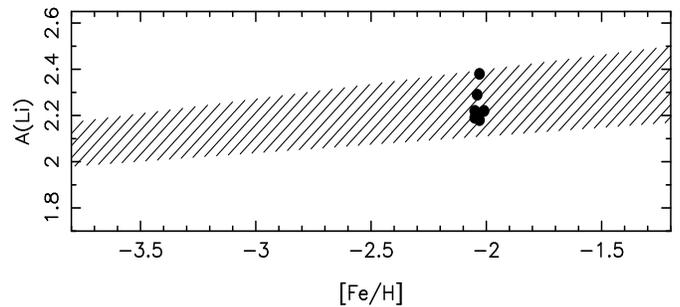}
\caption[]{Derived Li abundances of the program stars (filled dots)
compared with the diagram by Ryan et al.(2001) (shaded band). Error bars
are $\pm$0.1 dex for both axes.}
\end{figure}

It is important to note that lithium should not have been affected by dilution 
in the program stars; they have an effective temperature higher than the 
one needed to start the first dredge-up
(this happens around 5700K in a stellar model typical of our NGC 6397 sample;
see e.g. Charbonnel et al. 2000).
The mean Li abundance of our sample, in a scale
where $\varepsilon$(Li) = log~[n(Li)/n(H)]+ 12, is equal to
$\varepsilon(Li)$
= 2.23$\pm$0.07. Again, we emphasize the small scatter among the stars.
Our results agree with the study by Pasquini \& Molaro (1996) on three turnoff 
stars
of NGC 6397 which are located in the same temperature range.
They are also consistent with the general trend of lithium abundance in 
metal-poor stars (see Fig.~7).
However, a larger dispersion of $\varepsilon$(Li) was found in subgiant stars 
of M92 by Boesgaard et al. (1998), who favor differential depletion due to 
different rotation histories to explain their data. 
It is worth noticing that the M92 subgiants show other peculiar 
abundances, in particular under and overabundances of Mg and Al respectively 
(see \S 1) which could be due to intra-cluster pollution. 
More data for lithium are needed in globular cluster dwarfs. 
This is of particular importance if one wants to combine Li data of
cluster and halo stars to constrain the primordial abundance of this element
and its evolution in the early Galaxy.

\section{Conclusions}

We derived chemical abundances for several elements in seven turnoff stars in
NGC 6397. The cluster metallicity is [Fe/H] = -2.02$\pm$0.01 and
no significant star-to-star abundance variations were detected, indicating that 
NGC 6397 is a homogeneous cluster. Moreover, the iron abundance derived
from turnoff stars are in good agreement with abundances derived from subgiants
(Castilho et al. 2000) as expected.

Calcium and titanium have abundances in agreement with those derived for 
giants
and subgiants in this cluster, while the abundances of the iron peak elements 
agree with those
of field stars with similar iron content. These abundances are consistent
with the predictions of the stochastic model by Argast et al. (2000), when
the expected abundance ratio dispersion at a given metallicity is taken into
account, in spite of the uncertainties still present in the supernova yields.
We note that our [Ca/Fe] and [Mg/Fe] ratios match onto the lumpy 
structures found in the plots by Idiart \& Th\'evenin (2000), when halo dwarfs 
are considered. More data on clusters with different metallicities are
certainly required for a better understanding of these structures.

Magnesium is almost solar and small variations among the stars were
found within the expected errors,
but a definitive conclusion must wait for a larger sample. In
agreement with Gratton et al. (2000), no Mg vs Na anticorrelation seems to be
present in our turnoff stars, although an intriguing enhancement of the
magnesium abundance in subgiants, with respect to dwarfs, is suggested from
the data of these authors, again concerning only a few objects.

When data from different authors are considered (Minniti et al. 1996; Castilho
et al. 2000; Gratton et al. 2000), they suggest that
sodium variations are clearly present in evolved stars. 
The mean Na abundance ratio in dwarfs is almost solar, but there is
some evidence of intriguing star-to-star variations. A447 is deficient by
almost a factor of three with respect to the other stars and this trend is 
also verified in dwarfs studied by Gratton et al. (2000). Similar Na variations
in turnoff stars were also detected in 47 Tuc and in NGC 6752 (Cottrel \&
Da Costa 1981; Briley et al. 1994, 1995; Gratton et al. 2000), which 
are difficult to understand in either the framework of the
primordial or deep-mixing scenarios. The reality of these variations
must be confirmed by other studies and by analyses adopting the same procedure,
atmospheric models, and physical parameters to derive NLTE chemical abundances.
 
\acknowledgements {}
We thank the UVES team for building an excellent spectrograph and
the ESO staff at Paranal Observatory for the service observations.
We are grateful to the referee for his comments, which improved the
presentation of this paper and also to B. Gladman for other
improvement of the manuscript.
We thank INSU and PNPS for financial supports.
Part of this work has been performed using the computing facilities provided by 
the program
``Simulations Interactives et Visualisation en Astronomie et M\'ecanique 
(SIVAM)'' at the observatoire de la C\^ote d'Azur.

\end{document}